# Metallic liquid H$_3$O in a thin-shell zone inside Uranus and Neptune

Peihao Huang[1†], Hanyu Liu[1†], Jian Lv[1], Quan Li[1,4], Chunhong Long[2], Yanchao Wang[1*], Changfeng Chen[3*], Yanming Ma[1,4*]

[1]State Key Laboratory of Superhard Materials & Innovation Center for Computational Physics Methods and Software, College of Physics, Jilin University, Changchun 130012, China.
[2]Beijing Computational Science Research Center, Beijing 100193, China.
[3]Department of Physics and Astronomy, University of Nevada, Las Vegas, Nevada 89154, United States.
[4]International Center of Future Science, Jilin University, Changchun 130012, China.
†These authors contributed equally to this work

**The Solar System harbors deep unresolved mysteries despite centuries-long study. A highly intriguing case concerns anomalous non-dipolar and non-axisymmetric magnetic fields of Uranus and Neptune that have long eluded explanation by the prevailing theory. A thin-shell dynamo conjecture captures observed phenomena but leaves unexplained fundamental material basis and underlying mechanism. Here, we report the discovery of trihydrogen oxide (H$_3$O) in metallic liquid state stabilized at extreme pressure and temperature conditions inside these icy planets. Calculated stability pressure field compared to known pressure-radius relation for Uranus and Neptune places metallic liquid H$_3$O in a thin-shell zone near planetary cores. These findings from accurate quantum mechanical calculations rationalize the empirically conjectured thin-shell dynamo model and establish key physical benchmarks that are essential to elucidating the enigmatic magnetic-field anomaly of Uranus and Neptune, resolving a major mystery in planetary science.**

Planetary magnetic fields are commonly attributed to the magnetohydrodynamic dynamo. Earth's geomagnetic field is generated in a thick layer of fluid iron outer core, and the magnetic fields of Jupiter and Saturn are produced by their respective thick layers of metallic fluid hydrogen. Magnetic fields of most planets in the Solar System are axial-dipole dominated, except for Uranus and Neptune that exhibit anomalous non-dipolar and non-axisymmetric fields. This intriguing phenomenon raises fundamental questions about the physics mechanism and underlying material foundation.

Uranus and Neptune each contain three constituent regions[1], namely a gaseous hydrogen-helium atmosphere, a thick intermediate layer of "hot ices", and a rocky core. The pressure and temperature in their atmospheres are insufficient to produce metallic hydrogen[2,3], and iron in their cores is probably in solid state[3,4] and thus inconsequential in the dynamo model. The middle "ice layer" is therefore the most likely place to generate magnetic fields. This layer is estimated to possess 56% $H_2O$, 36% $CH_4$, and 8% $NH_3$ in molar fractions[5], and its pressure and temperature conditions range from 20 to 600 GPa and 2,000 to 7,000 K, respectively. The $H_2O$ in the form of superionic ice may constitute a large fraction of the ice layer[5-8] and ionic fluid water exists in the remaining outer part of the ice layer. The conductivity of the ionic ice, estimated to be 2 to 20 $(\Omega\ cm)^{-1}$, is too small to generate the magnetic fields in Uranus and Neptune[9,10]. The conductivity of the superionic ice is near 100 $(\Omega\ cm)^{-1}$, which is still too small and need to be compensated by vigorous convection with flow velocities similar to those stirring Earth's liquid outer core (~100 m $yr^{-1}$) to sustain a dynamo in a superionic ice layer, but the requirements for this scenario are considered unrealistic[7].

It is known that $CH_4$ becomes unstable and dissociates into C and $H_2$ above 10 GPa and 2,000 K[11-14], and $NH_3$ reacts with water to produce molecular mixtures in the form of $(H_2O)(NH_3)_2$[15] from 79 to 540 GPa. Given the estimated molar fractions of $H_2O$, $CH_4$, and $NH_3$ in their ice layers[5], the mantle regions of these planets are expected to host a hydrogen-rich environment with the $H_2O$:$H_2$ ratio reaching 13:18 assuming full dissociation of $CH_4$ and complete mixing of $NH_3$ with $H_2O$ in a ratio of 2:1.

Past studies examined miscibility of $H_2$ and $H_2O$ in pressure and temperature ranges of 2-70 GPa and 1,000-6,000 K, respectively,[16] which do not correspond to conditions deep inside Uranus and Neptune. It remains a viable prospect that $H_2$ could react with $H_2O$, which is the most abundant component in the ice layer of these planets, to produce new compounds that may provide the material basis for generating magnetic fields.

We examined $H_2$-$H_2O$ systems using an inhouse material structure search method[17,18]. Our search uncovered trihydrogen oxide ($H_3O$) that is stable at 600 GPa (Fig. 1a). This stoichiometry was previously reported but at much higher pressures (~14 TPa)[19]. The unit cell of $H_3O$ contains 64 atoms, and its hydrogen-oxygen covalent framework has a H:O ratio of 2:1, with remaining hydrogen in the voids in the form of $H_2$ molecules. The basic structural unit hosts six distinct bond lengths and three different bond angles (Fig. 1c). Adjacent basic units are nested, and the unit cell contains 4 basic units. The bond length in the $H_2$ unit is 0.65 Å at 500 GPa, which is much shorter than that in solid hydrogen at ambient pressure (0.74 Å). Further assessments of formation enthalpy show that $H_3O$ is stable against decomposition to $H_2O$ and $H_2$ above 450 GPa (Fig. 1b).

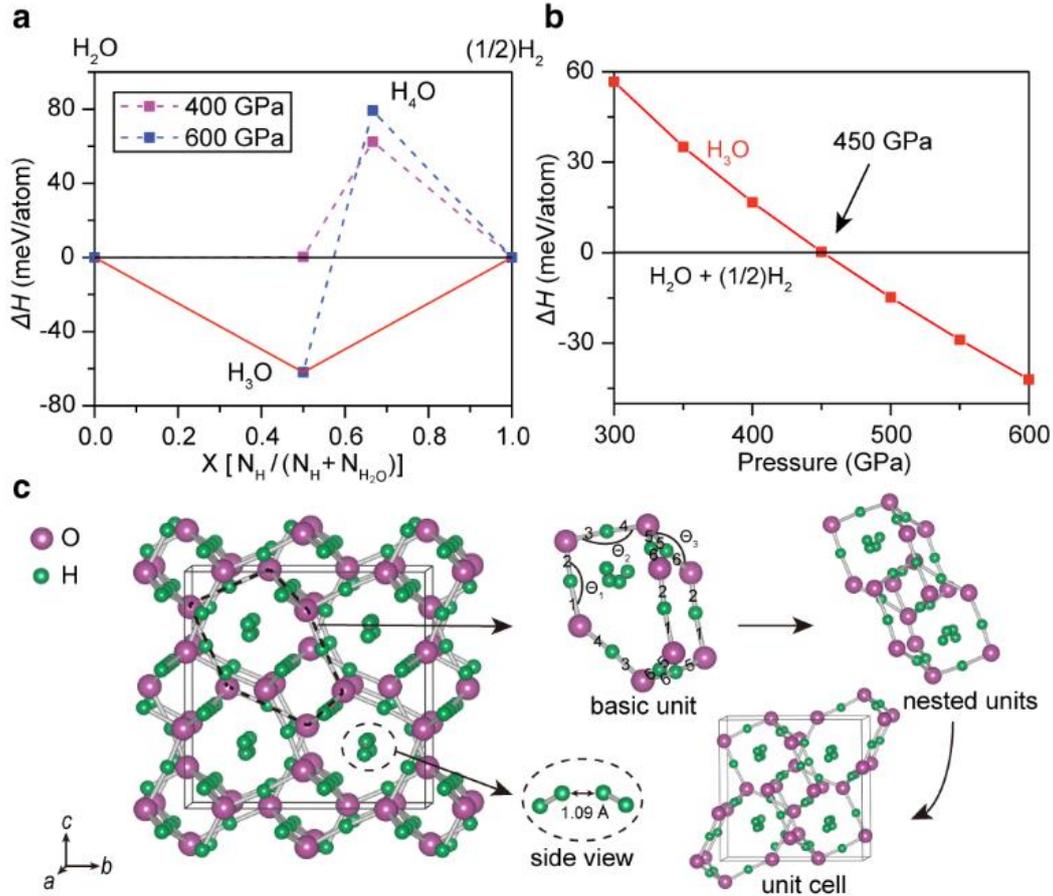

**Fig. 1 | Formation enthalpy and crystal structure of $H_3O$. a**, Formation enthalpy of H-O compounds with respect to decomposition into $H_2O$ and $H_2$ at 400 and 600 GPa. **b**, Enthalpy including the effect of zero-point energy versus pressure for $H_3O$ relative to the results for decomposed $H_2O$ and $H_2$. **c**, Crystal structure (space group *Cmca*) and basic structural units of $H_3O$. Lattice parameters at 500 GPa are a=3.35 Å, b=5.84 Å, and c=5.80 Å. There are two inequivalent atomic positions for O and four for H: O1 at 8f (0.00, 0.16, 0.00) and O2 at 8f (0.50, 0.49, 0.34), H1 at 8f (0.00, 0.33, 0.07), H2 at 8f (0.00, 0.43, 0.34), H3 at 16g (0.76, 0.09, 0.09) and H4 at 16g (0.33, 0.24, 0.23). Six distinct bonds in the basic unit labeled 1 through 6 have lengths 1.16, 1.02, 1.08, 1.06, 1.06 and 1.10 Å, respectively, and three bond angles $\theta_1$=174.58°, $\theta_2$=174.67° and $\theta_3$=171.33°. Adjacent basic units are nested to form the unit cell of the $H_3O$ crystal.

The stability pressure field of 450-600 GPa for $H_3O$ lies within the pressure range of 20-600 GPa in the interiors of Uranus and Neptune, which also host high temperatures. We examined pertinent pressure-temperature phase diagram using *ab initio* molecular dynamics (AIMD) simulations. Fig. 2 shows simulated mean square displacement (MSD) and trajectory of $H_3O$ at the density of 4.30 g/cm$^3$. At 1,000 K, MSD values of hydrogen and oxygen atoms remain nearly constant (Fig. 2a), showing the system in a stable solid state with all atoms near equilibrium positions (Fig. 2d). At 3,000 K, MSD values for hydrogen atoms begin to increase, but values for oxygen atoms remain nearly constant (Fig. 2b), signaling a transition into superionic state, corroborated by trajectory data (Fig. 2e) showing that oxygen atoms remain near their equilibrium positions while hydrogen atoms diffuse. At further increased temperature (6,000 K), oxygen atoms also diffuse (Fig. 2c and 2f). Systematic calculations show that $H_3O$ transforms from solid to superionic state at 1,250 K at 4.30 g/cm$^3$, then to liquid state at 5,250 K (see Fig. 3).

We also ran simulations for $H_2O$ at the density of 4.93 g/cm$^3$ and temperatures of 5,000-8,000 K, with the equilibrium pressure (~600 GPa) close to that for $H_3O$ at 4.30 g/cm$^3$ and 7,000 K. We find that $H_2O$ remains in a superionic state up to 7,000 K, where $H_3O$ enters a liquid state. We performed AIMD simulations using a supercell (Extended Data Fig. 1a) containing 144 atoms in the *Pbcm*-$H_2O$[20] structure and 48 atoms in the *I*4$_1$/*amd*-H structure[21] in the simulation box at 4.30 g/cm$^3$ and 7,000 K. The results show that hydrogen atoms can easily penetrate into the $H_2O$ lattice (Extended Data Fig. 1b), and the radial distribution function of the H-$H_2O$ mix matches that of $H_3O$ (Extended Data Fig. 1c). These results validate the scenario of $H_2O$ reacting with H to form liquid $H_3O$ at such extreme planetary pressure and temperature conditions.

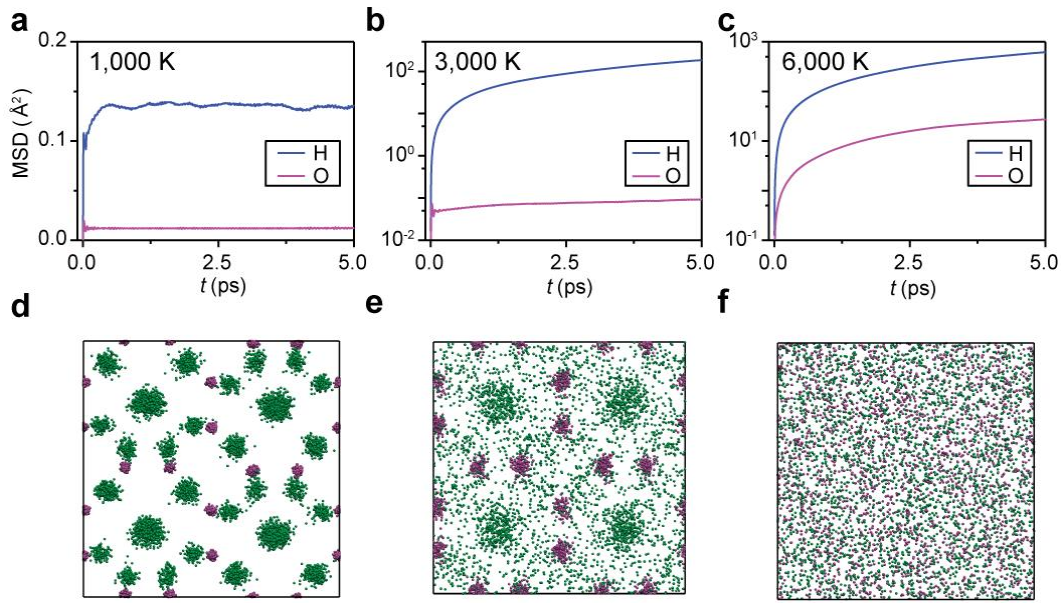

**Fig. 2 | Simulation of solid, superionic and liquid phase of $H_3O$. a-c**, Mean square displacement of oxygen and hydrogen in $H_3O$ at (**a**) 1,000 K, (**b**) 3,000 K, and (**c**) 6,000 K at the density of 4.30 g/cm$^3$, in solid, superionic and liquid phases. **d-f**, Trajectory snapshots of the corresponding phases of $H_3O$, where purple and green spheres represent instantaneous positions of oxygen and hydrogen atoms, respectively.

We have constructed the pressure-temperature phase diagram of $H_3O$ covering the conditions corresponding to the interiors of Uranus and Neptune (Fig. 3a). Phase boundaries (dashed lines) separating the solid, superionic and liquid states are determined by AIMD simulations. The boundary separating $H_3O$ and its decomposition compounds ($H_2O$ and $H_2$) at low temperatures (red open symbols) is determined by the difference of Gibbs free energy estimated using the quasi-harmonic approximation[22]. The isentropes of Uranus and Neptune pass through regions of the liquid phase of $H_3O$ above ~518 GPa, indicating that liquid $H_3O$ is viable deep inside these icy giants surrounding their rocky cores. From an estimated pressure-radius relationship for these planets[23], liquid $H_3O$ can stably exist in a thin-shell zone between 0.32 R to 0.38 R from the planetary center (Fig. 3b), where R is the total radius of the planet.

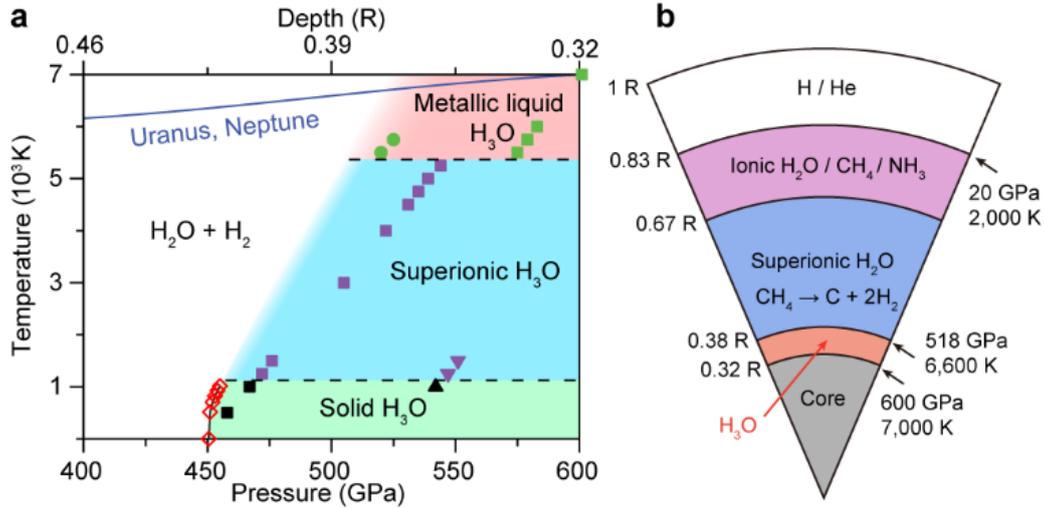

**Fig. 3 | The pressure-temperature phase diagram of $H_3O$ and its presence in a thin-shell zone inside Uranus and Neptune. a**, Stability fields separating $H_2O+H_2$ and solid, superionic, and metallic liquid phases of $H_3O$, obtained from AIMD simulations. The blue solid line is the isentrope of Neptune or Uranus[24]. Data points in filled circles, filled squares and filled triangles are AIMD simulation results at 4.14, 4.30 and 4.52 g/cm$^3$, respectively. **b**, Schematic of planetary interior structure of Uranus and Neptune. The grey and white regions represent the core and outer gas layer, respectively. The purple, blue and orange areas represent regions dominated by ionic $H_2O$, superionic $H_2O$ and liquid $H_3O$, respectively. The pressure-radius relation is extracted from the estimate by Ravit Helled *et al.*[23].

Electrical conductivity is crucial to understanding planetary magnetic fields in the dynamo theory. Our electronic bandgap calculations show that $H_3O$ has much lower metallization temperature (~5,000 K at 539 GPa) compared to $H_2O$ (~6,000 K at 587 GPa) (Extended Data Fig. 2a). This result indicates that $H_3O$ turns into a metallic liquid in deep mantle regions of Uranus and Neptune; meanwhile, $H_2O$ remains superionic and does not turn into a metallic liquid at the highest temperature and pressure conditions (7,000 K and 600 GPa) in the ice layer of these planets. Calculated direct current (DC) electrical conductivity results (Extended Data Fig. 2b) show an abrupt increase of conductivity between 5,000 and 6,000 K for $H_3O$, from 19 to 164 $(\Omega cm)^{-1}$, which is consistent with the bandgap results. Meanwhile, electrical conductivity of $H_2O$ shows no sudden change, and is an order of magnitude lower than that of $H_3O$ above 6,000 K (Extended Data Fig. 2b). The metallization of $H_3O$ may be attributed to a similar mechanism underlying melting-prompted conduction in diamond[25] and $N_2$[26].

A conjectured new planetary structural model[27,28] is able to reproduce the unusual magnetic fields of Uranus and Neptune in the context of the dynamo theory. The central component of this model is a thin shell of electrically conducting fluid, in contrast to thick rotating shells in other planets possessing magnetic fields. The present study predicts a narrow radial range for $H_3O$ phase stability in deep mantle regions of Uranus and Neptune, providing compelling evidence for the conjectured thin-shell structure[27,28]. Moreover, fluids with high electrical conductivity are essential to generating planetary magnetic fields[29]. Our calculations reveal that electrical conductivity of $H_3O$ reaches 164 $(\Omega cm)^{-1}$ at 6,000 K, an order of magnitude larger than the conductivity of ~20 $(\Omega cm)^{-1}$ for ionic ice, which was previously proposed to be responsible for the magnetic field of Uranus[30]. Liquid $H_3O$ is also more prone to convection than superionic ice, whose viscosity is several orders of magnitude larger than that of typical fluids, making $H_3O$ more conducive to producing magnetic fields. The present findings provide crucial foundation for resolving the magnetic-field anomaly of Uranus and Neptune, and the insights may prove useful in exploring pertinent phenomena in giant icy exo-planets.

**Acknowledgements** We acknowledge funding from the National Natural Science Foundation of China (Grant No. 11822404, 11774127, 11534003 and 11622432), the National Key Research and Development Program of China under Grant No. 2016YFB0201200, No. 2016YFB0201201, and No. 2017YFB0701503, Science Challenge Project No. TZ2016001, and Program for JLU Science and Technology Innovative Research Team (JLUSTIRT). We utilized computing facilities at the High Performance Computing Center of Jilin University and Tianhe2-JK at the Beijing Computational Science Research Center.

**Author Contributions** Y.M. and Y.W. conceived and designed the project. Y.W. predicted the $H_3O$ structure. P.H. performed simulations. P.H., H.L., Y.W., Y. M., C.C., J.L. and Q.L. performed data analysis. All authors wrote the manuscript and contributed to discussions of the results and revision of the manuscript.

**Competing Interests** The authors declare no competing interests.

**Materials & Correspondence** Correspondence and requests for materials should be addressed to Y.W. (wyc@calypso.cn), C.C. (chen@physics.unlv.edu), and Y.M. (mym@calypso.cn; mym@jlu.edu.cn).

**Methods**

We have performed extensive structure searches through a swarm-intelligence-based CALYPSO method and its same-name code, which enables global structure searching in conjunction with *ab initio* energetic calculations. This method has been benchmarked on various known systems.[31-34] The DFT calculations are performed using the VASP plane-wave code[35,36], where the Perdew-Burke-Ernzerhof (PBE) generalized gradient approximation density functional[37] and frozen-core all-electron projector-augmented wave (PAW) potentials[38] are adopted. The cutoff radius of the pseudopotentials for oxygen and hydrogen are 0.58 Å and 0.42 Å, respectively. To ensure validity of the adopted pseudopotentials, we also performed full-potential all-electron calculations for the equation of states of $H_3O$ over the pressure range studied here by using the WIEN2K code[39]. The results of our VASP calculations are nearly identical to those obtained from all-electron calculations (Extended Data Fig. 3), validating the pseudopotentials up to 600 GPa. The electronic wave functions are expanded in a plane-wave basis set with a kinetic energy cutoff of 1,200 eV, and the Brillouin zone (BZ) sampling is performed on k-meshes with a reciprocal space resolution of $2\pi \times 0.032$ Å$^{-1}$ to ensure that energies are converged to 1 meV/atom. The phonon calculations are performed through the PHONOPY code[40] using a finite displacement approach[41].

The AIMD simulations are performed in the canonical (NVT) ensemble applying a Nosé thermostat. A system containing 128 atoms is used for $H_3O$. Simulations reach 15 ps, with the first 2-5 ps for equilibrating the system. The time step is chosen to be 0.5 fs. A 2×2×2 k-mesh grid is used for the BZ sampling. To check for size effect, a larger system with 576 atoms is adopted. The transition temperature of this larger system is the same as that of the smaller system, confirming convergence regarding system size. The simulated temperatures range from 1,000 to 7,000 K with a temperature step of 500 K, and we choose the temperature step in interested regions (1,000-1,500 K and 5,000-6,000 K) to be 250 K to investigate the phase-transition temperatures from solid to superionic phase, and then to liquid phase. The simulation cell of $H_2O$ contains 144 atoms. The Gibbs free energy (G) is calculated via quasi-harmonic approximation by the PHONOPY code. Gibbs free energy is defined at a constant temperature (T) and pressure (P) by the formula: $G(T, P) = \min_{V}[U(V) + PV + F_{phonon}(T, V)]$, where U is the internal lattice energy, V is volume and $F_{phonon}$ is the phonon free energy. $F_{phonon}(T, V) = k_B T \int_0^\infty g(\omega, V) \ln[2\sinh(\frac{\hbar\omega}{2k_B T})]d\omega$, where $g(\omega, V)$ is the phonon density of states at frequency ω and volume V. The minimal value of G is found at the equilibrium volume for a given T and P.

The time-averaged bandgap is calculated by the PBE functional (50 configurations for averaging) and cross checked using the GW methods (10 configurations for averaging). For GW calculations, a 2×2×2 k-mesh grid is used for balancing accuracy and efficiency (Extended Data Fig. 4). Calculations show that the PBE and GW results are in good accordance, with consistent conclusion about $H_3O$ reaching metallic state (Extended Data Fig. 5) under the same pressure and temperature conditions. The time-averaged DC electrical conductivity is calculated via the Kubo-greenwood formula[42,43] using the KGEC code[44], which is a post-processor module for use with Quantum Espresso[45]. To obtain converged conductivity results, 576 atoms for $H_3O$ and 540 atoms for $H_2O$ are used. A 3×3×3 k-mesh is used for BZ sampling. Each conductivity result is evaluated by averaging over 10 configurations.

**Data availability**

The data that support the findings of this study are available from the corresponding author upon reasonable requests.

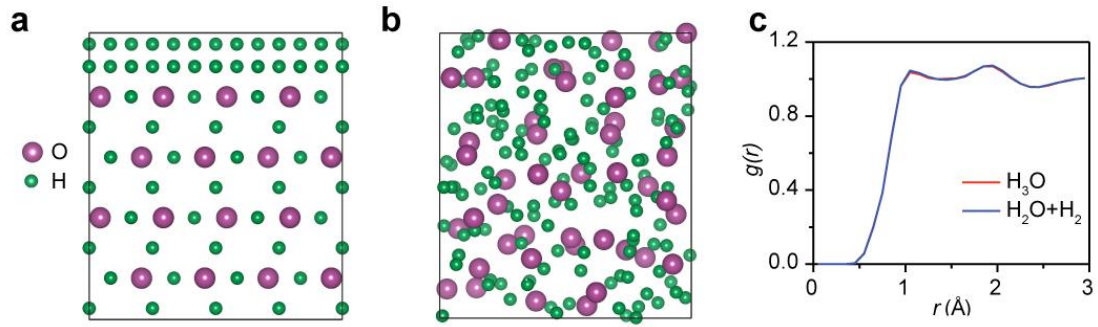

**Extended Data Fig. 1 | Simulation of mixing 144 atoms of $H_2O$ and 48 atoms of $H_2$ in a supercell. a**, Initial construction of the supercell with the $H_2O$ and $H_2$ components. **b**, A snapshot of the last step of the AIMD simulation at 4.30 g/cm$^3$ and 7,000 K. The density is chosen so that simulated pressures lie well inside the range of the ice layer in Uranus and Neptune. **c**, Comparison of radial distribution function of the equilibrated $H_2$-$H_2O$ mixing and the corresponding result for the $H_3O$ phase described in the main text. The two sets of results appear nearly identical, highlighting the main conclusion that superionic $H_2O$ reacts with $H_2$ to form $H_3O$ at the conditions of deep-mantle ice layers of Uranus and Neptune.

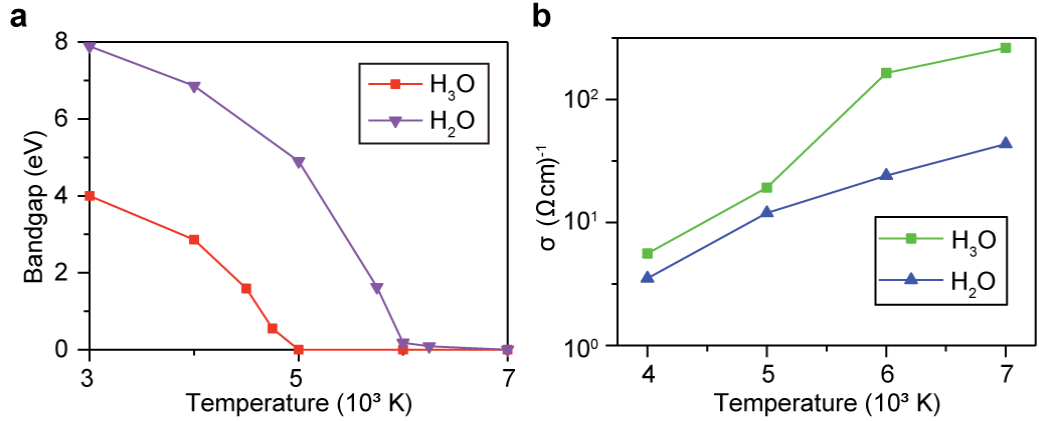

**Extended Data Fig. 2 | Metallicity of liquid H₃O. a**, Evolution of time-averaged bandgap for H$_3$O (at the density of 4.30 g/cm$^3$, with an equilibrium pressure ~600 GPa at 7,000 K) compared with H$_2$O (at the density of 4.93 g/cm$^3$, with an equilibrium pressure ~600 GPa at 7,000 K) as a function of temperature obtained by using the GW method that is known to give reliable bandgap values. **b**, Calculated time-averaged direct-current (DC) electrical conductivity for H$_3$O compared with the results for H$_2$O as a function of temperature obtained using the Kubo-greenwood formula.

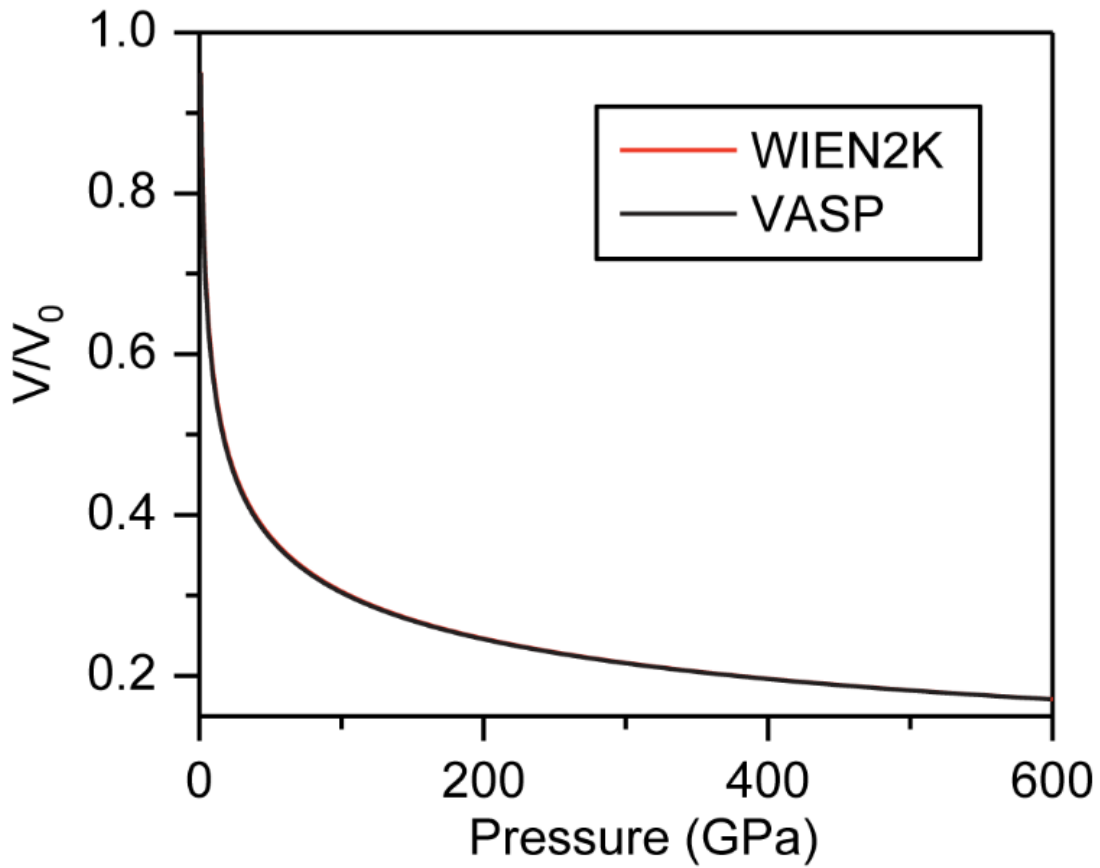

**Extended Data Fig. 3 | Equation of states of $H_3O$ calculated by VASP and WIEN2K.** To test the adequacy of the pseudopotentials used in the calculations at high pressures examined in the present work, we have computed the equation of states of $H_3O$ by using the VASP code, which adopts the pseudopotentials, and compared the results against those obtained using the full potential as implemented in the WIEN2K code. Calculated results presented here show that these two computational methods produce nearly identical data, thus validating the pseudopotentials adopted in the VASP calculations over the wide pressure-volume range examined in the present work.

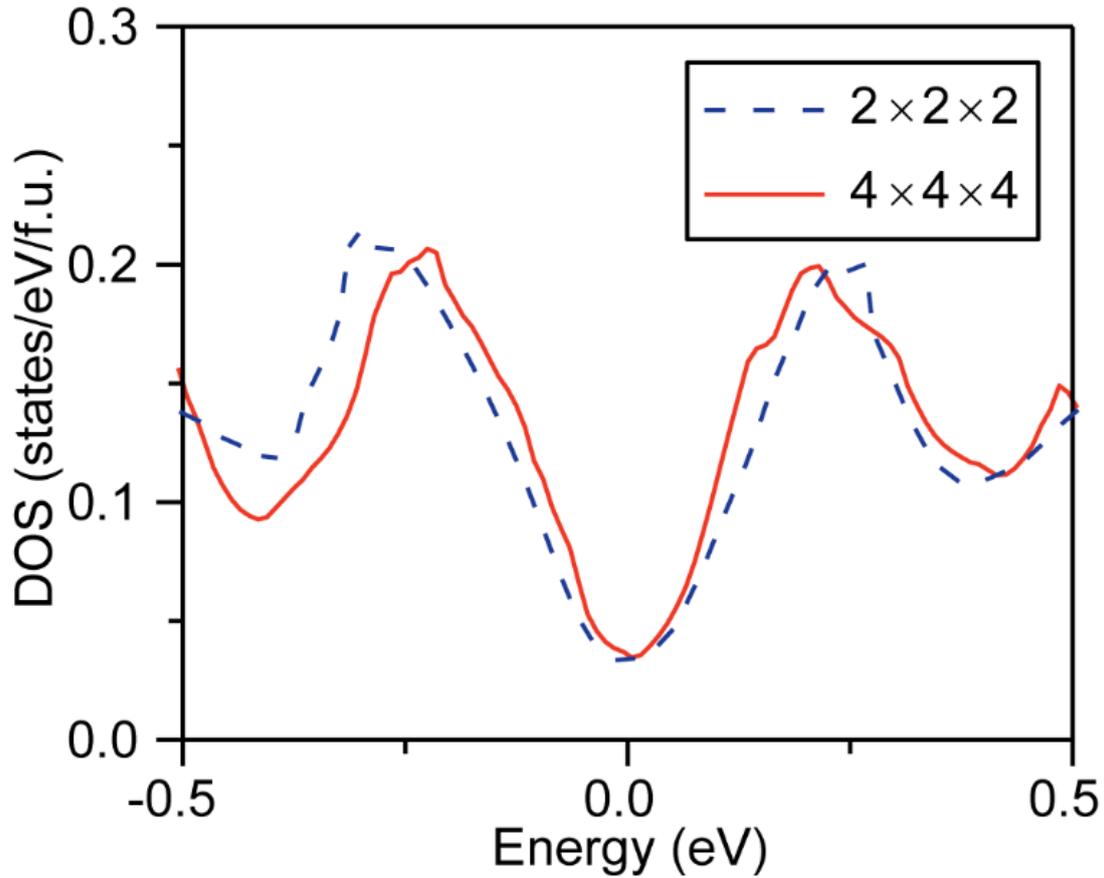

**Extended Data Fig. 4 | Convergency test of the time-averaged electronic density of states for H$_3$O with respect to the k-mesh grid used in the GW calculations.** To balance computational efficiency and reliability, we have performed convergency test on k-mesh adopted in the calculations of electronic density of states reported in the present work. The calculations were performed at the density of 4.30 g/cm$^3$, with an equilibrium pressure ~600 GPa at 7,000 K. A comparison of the computed data obtained using a 2×2×2 and a 4×4×4 mesh indicates that the former set of results is well converged, and therefore the 2×2×2 mesh is employed in the calculations of the time-averaged electronic density of states presented in the main text.

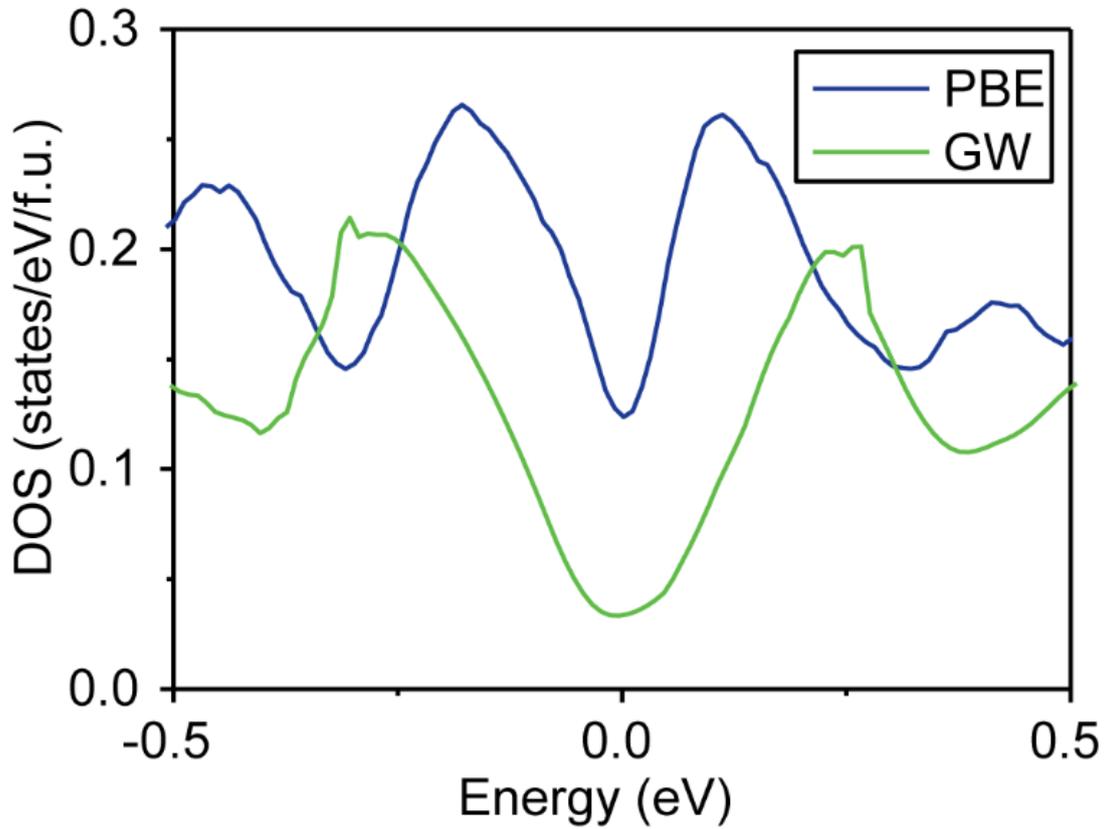

**Extended Data Fig. 5 | The time-averaged electronic density of states of H$_3$O at 7,000 K by the PBE-GGA functional and the GW method.** Two sets of calculations were performed to examine the reliability and consistency of the results obtained from computations using the PBE functional and the GW method. The calculations were performed at the density of 4.30 g/cm$^3$, with an equilibrium pressure ~600 GPa at 7,000 K. Results presented here show that both methods produce consistent outcome indicating that H$_3$O is metallic at 7,000 K, despite some quantitative differences in charge distribution, which have no impact on our main conclusions about the key benchmarks on metallicity of H$_3$O under the physical conditions studied here.